\RequirePackage{lineno}
\pdfoutput=1

\documentclass[prl,twocolumn,nofootinbib,floatfix,subeqn]{revtex4}
\usepackage{graphicx}
\usepackage{amsmath}
\usepackage[english]{babel}
\usepackage[short]{datetime}

\newcommand{\Ui}{U_\mathrm{int}}
\newcommand{\Erec}{E_\mathrm{r}}
\newcommand{\llat}{\lambda_\mathrm{lat}}
\newcommand{\Vlat}{V_\mathrm{lat}}
\newcommand{\Vt}{V_\mathrm{t}}
\newcommand{\alat}{a_\mathrm{lat}}
\newcommand{\mrm}[2]{#1_\mathrm{#2}}
\newcommand{\ket}[1]{\left|#1\right>}
\newcommand{\bra}[1]{\left<#1\right|}
\newcommand{\Rb}{^{87}\mathrm{Rb}}

\newcommand{\Tr}{T_\mathrm{r}}

\begin{document}



\title{A quantum computation architecture using optical tweezers}

\author{Christof Weitenberg$^{1}$}
\altaffiliation{present address: Laboratoire Kastler Brossel, Ecole Normale Sup\'{e}rieure, 24 rue Lhomond, F-75005 Paris, France}
\author{Stefan Kuhr$^{1,2}$}
\author{Klaus M\o lmer$^{3}$}
\author{Jacob F. Sherson$^{3}$}\email{sherson@phys.au.dk}

\date{\today \ \  \currenttime}

\affiliation{
   $^1$Max-Planck-Institut f\"ur Quantenoptik, Hans-Kopfermann-Str.~1, 85748 Garching,
   Germany\\
   $^2$University of Strathclyde, Department of Physics, SUPA, Glasgow G4 0NG, United Kingdom\\
   $^3$Department of Physics and Astronomy, University of Aarhus, DK-8000 Aarhus C, Denmark
}

\begin{abstract}
We present a complete architecture for scalable quantum computation with ultracold atoms in optical lattices using optical tweezers focused to the size of a lattice spacing.
We discuss three different two-qubit gates based on local collisional interactions. The gates between arbitrary qubits require the transport of atoms to neighboring sites.
We numerically optimize the non-adiabatic transport of the atoms through the lattice and the intensity ramps of the
optical tweezer in order to maximize the gate fidelities. We find overall gate times of a few $100\,\mu$s, while keeping the error probability due to vibrational excitations and spontaneous scattering below $10^{-3}$.
The requirements on the positioning error and intensity noise of the optical tweezer and the magnetic field stability are analyzed and we show that atoms in optical lattices could meet the requirements for fault-tolerant scalable quantum computing.
\end{abstract}

\maketitle



\section*{I. Introduction}

Neutral atoms in periodic optical potentials have long been considered a promising candidate for scalable quantum computation due to long coherence times of internal state qubits and their excellent controllability. However, a great challenge in these systems remains to tailor interactions for the two-qubit gates.
One approach is to use dipole-dipole interactions~\cite{Brennen:1999}, most prominently among Rydberg atoms~\cite{Jaksch:2000}, realized recently with individual atoms in separate dipole traps~\cite{Isenhower:2010,Wilk:2010}.
A different approach uses ground state collisions either within the same trapping potential~\cite{Jaksch:1999,Calarco:2000,Briegel:2000,Negretti:2005} or mediated by tunneling~\cite{Pachos:2003,Strauch:2008}. In optical lattices, two-qubit gates have been implemented on many pairs of atoms in parallel~\cite{Mandel:2003,Anderlini:2007} but they have, so far, not been realized on a single pair of atoms.
The main challenge  is to attain sufficient resolution to manipulate single lattice sites. Several proposals to overcome the diffraction limit have been made to achieve this goal~\cite{You:2000,Schrader:2004,Saffman:2004,Cho:2007}, and recently single-site manipulation by an optical tweezer  in a short-period optical lattice has been realized~\cite{Weitenberg:2011}  and could now be exploited for the implementation of single- and two-qubit gates.

In this work, we propose and investigate an architecture for scalable quantum computation in optical lattices. Using tightly focussed optical tweezers,
atomic qubits are transported around in the lattice and merged into single lattice sites in order to implement collisional quantum gates.
We adapt the $\sqrt{\rm swap}$ gate, proposed in Ref.\,\cite{Hayes:2007} and realized in Ref.\,\cite{Anderlini:2007} to our tweezer-based optical-lattice architecture. By
moving the atoms non-adiabatically, our transport and two-bit gate times become much faster than, \textit{e.g.}, the tunneling gate proposed in Ref.\,\cite{Pachos:2003}. Similar to work in Ref.\,\cite{DeChiara:2008}, we numerically optimize the ramp-up of the tweezer intensity and the transport, but due to the difference in the available control parameters, we obtain smaller errors, and arrive at total gate times of a few hundred microseconds with error probability $10^{-3}$ arising from non-adiabatic excitations and spontaneous scattering. This scheme can in principle allow for thousands of operations within the coherence time of the qubit.
We also investigate the requirements for the stability of the position and intensity of the tweezer as well as the effect of magnetic field noise.

\begin{figure}[b]
  \centering
  \includegraphics[width=0.9\columnwidth]{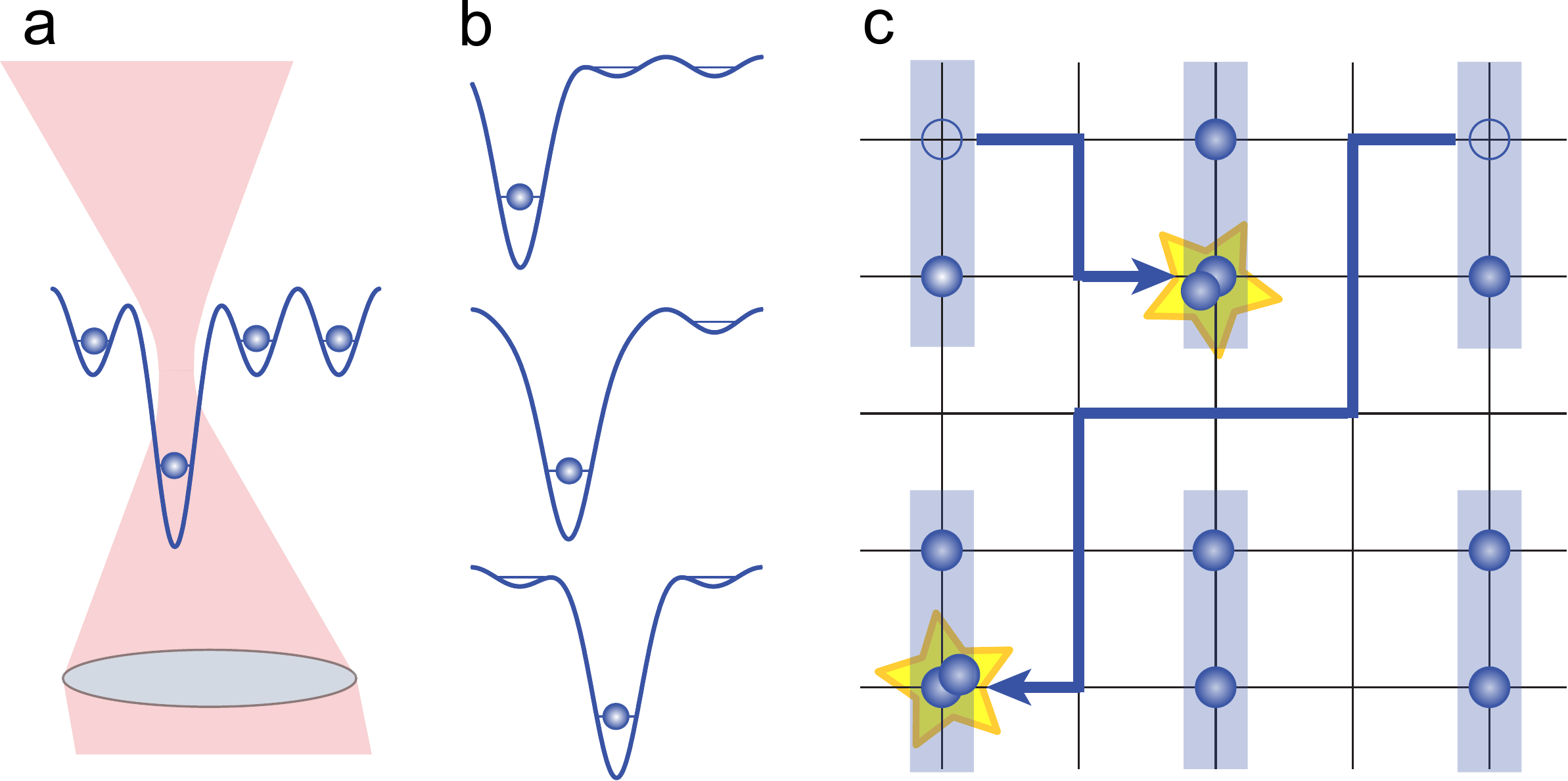} 
  \caption{
  Optical tweezer-based architecture.
  a) A tightly focused optical tweezer locally perturbs the optical lattice potential.
  b) The atoms can be shuttled around by moving the tweezer.
  c) The quantum register is initialized with an atomic pattern with empty rows and columns to permit unhindered transport of atoms within the lattice.
  The two-qubit gates utilize the collisional interactions between atoms positioned at the same site. With multiple tweezers in use, several gates can be performed in parallel.}
  \label{fig:setup}
\end{figure}

The paper is organized as follows. Section\,II introduces the scalable architecture, including the quantum register initialization, the single-qubit gates and the readout. Section\,III investigates the non-adiabatic transport of atoms through the lattice, which is required for the collisional two-qubit gates. In Section\,IV, we present three different gates and discuss their respective advantages including the role of coherence times of the qubit states used. In section\,V, the different errors introduced by spontaneous scattering and
by fluctuations of tweezer position and intensity are discussed and compared to the threshold for fault-tolerant quantum computing.


\section*{II. Scalable architecture}
In this section, we present our scalable architecture for quantum computing using optical tweezers for addressing and transport in optical lattices. The differential light shift of a tweezer on selected atoms can be exploited for single-qubit gates (Section\,II.C), while moving tweezers can bring arbitrary pairs of atoms together for collisional two-qubit gates.
Single-qubit rotations form a universal set of gates in combination with either the two-qubit phase gate~\cite{Mandel:2003} or the $\sqrt{\rm swap}$ gate~\cite{Loss:1998}. Together with the initialization (Section\,II.B) and the readout capability (Section\,II.D), the system therefore fulfills the requirements for scalable quantum computation~\cite{DiVincenzo:2000}.



\subsection*{A. General setup}
Throughout this paper we assume a 3D optical lattice of $\Rb$ atoms formed by three pairs of counter-propagating near-infrared laser beams (wavelength $\llat=1064\,$nm),  realizing a lattice with spacing  $\alat=\llat/2$. We assume the ability to prepare a single 2D system and the availability of an imaging system with a resolution such that all individual sites can be detected~\cite{bakr:2010,sherson:2010}
and individually addressed~\cite{Weitenberg:2011,Zhang:2006}. We will assume a lattice depth in all directions of $\Vlat=50\,\Erec$, where $\Erec=h^2/(2m\llat)$ is the recoil energy and $m$ is the mass of a $\Rb$ atom. This yields a trapping frequency $\omega_{\mathrm{trap}}\approx 2\pi\cdot30\,\mathrm{kHz}$, a two-particle interaction energy $\Ui/h\approx2\,\mathrm{kHz}$, and a tunnel coupling $J/h\approx0.06\,\mathrm{Hz}$.

The qubit will be composed of two ground state hyperfine levels of $\Rb$ but unlike in Ref.~\cite{Weitenberg:2011} we will assume the atoms to be addressed by a tweezer close to the 5S-6P transition at $\lambda_{1/2}=421.555 ~\mathrm{nm}$ and  $\lambda_{3/2}=420.1733 ~\mathrm{nm}$. Tuning the wavelength between these two transitions yields a differential light shift which we will exploit for spin-dependent transport.
We describe the optical tweezer by a Gaussian beam profile with a $1/e^2$ waist radius of $w_o=\alat/2$, which can be reached with an imaging system of numerical aperture 0.68.
Unless otherwise stated the maximum light shift of the optical tweezer is $\Vt=500\,\Erec$.
For the gate in Section\,IV.C, two tweezers are needed, and many tweezers are required for the parallelization of gates. These can be created and quickly controlled using acousto-optic deflectors~\cite{Zimmermann:2011}. 

\subsection*{B. Quantum register initialization}
A major advantage of ultracold atoms in optical lattices over other systems lies in the possibility of creating large scale quantum registers. This can be done by preparing the system in a Mott insulating state~\cite{Greiner:2002,bakr:2010,sherson:2010}, which pins the occupation of the lattice sites to integer values due to the on-site interaction between atoms and can thus realize unity occupation.
When the lattice is sufficiently deep, deviations from unity filling are only thermally activated and depend on the position in the external confinement. In the center of the Mott insulating domain, the deviations can be very small and current experiments reach $P_{n\neq 1}\approx 3\cdot10^{-3}$~\cite{sherson:2010}.
Different purification schemes have been proposed to circumvent residual thermal defects. They involve either illumination and subsequent recooling of all atoms~\cite{Weiss:2004}, many-body dynamics~\cite{Popp:2006,Ho:2009,Bernier:2009,Doria2011}, or algorithmic cooling schemes~\cite{Sherson:2010b}. Recent experiments have demonstrated a number-selective removal of atoms with an efficiency of 86\%\,\cite{Bakr:2011}.


Starting from the Mott insulator with unity filling, we remove every second column and additionally one row so that they can be used as a channel for shuttling atoms between different sites [Fig.\,\ref{fig:setup}c)]. The removal can be performed by transferring selected atoms to a different hyperfine ground state and subsequently removing them with a resonant laser illuminating the entire sample~\cite{Weitenberg:2011}.

\subsection*{C. Single-qubit gates}
A single-qubit gate can be realized using either microwave radiation or optical two-photon Raman transitions between two selected ground state hyperfine levels.
In the former case, the differential light shift of a tightly focused tweezer or a magnetic field gradient shifts selected atoms into resonance with a global microwave field driving the transition. In the latter case, the rotation is driven by two tightly focussed laser beams in a Raman  configuration with zero two-photon detuning.

Raman-based gates have been demonstrated in single optical dipole traps and reached a $\pi/2$ gate time of 183\,ns\,\cite{Yavuz:2006} and 37\,ns\,\cite{Jones:2007} with a next-neighbor residual rotation at the level of $10^{-3}$\,\cite{Yavuz:2006}. A theoretical analysis for the performance in a large spacing optical lattice yields an achievable error of $10^{-4}$ for Rb~\cite{Saffman:2005} and $10^{-5}$ for Cs~\cite{Beals:2008}.

Microwave-based gates were realized in a 1D lattice using a magnetic field gradient, reaching a $\pi/2$ gate time of $8\,\mu$s~\cite{Schrader:2004}.
In 2D, single spin manipulation was demonstrated using   the local light shift due to a focused laser beam and Landau-Zener microwave sweeps~\cite{Weitenberg:2011}.
A theoretical analysis suggests that gate times of $100\,\mu$s can be reached with an error probability of ending in the wrong state below $10^{-4}$ at a lattice spacing of $10\,\mu$m\,\cite{Beals:2008} and with an error $<10^{-3}$ at a spacing of $425\,$nm\,\cite{Zhang:2006}.
Recently an error of $1.4 \cdot 10^{-4}$ per global microwave-based single-qubit gate of $30\,\mu$s gate time was demonstrated for Rb atoms in an optical lattice\,\cite{Olmschenk:2010} using randomized benchmarking~\cite{Knill:2008}.

Comparing the approaches, it is obvious that the Raman-based gates are much faster. However, since even the slower microwave-based gates are comparable with the two-qubit gate time discussed in Section\,IV   faster single-qubit rotations will not significantly speed up the overall computation.
The drawback of the Raman-based gate is that is extremely difficult to avoid cross-talk at the small lattice spacings envisaged here. On the other hand, the microwave-based gate  restricts the choice of the qubit states to magnetic field-sensitive states, which have shorter coherence times, but which are anyway required for some two-qubit gates, such as the one discussed in Section\,IV.B.



\subsection*{D. Readout}
As the final step, the spin state of several qubits has to be read out. For this, we can transfer one of the two qubit spin states into the neighboring free shuttling areas using the spin-dependent transport discussed in Section\,IV.B, followed by spatially resolved fluorescence imaging~\cite{Nelson:2007,bakr:2009,sherson:2010}. In the case of magnetic field-insensitive qubit states, one can remove one of the spin states and detect the qubit state by the presence or absence of an atom in the fluorescence image.

\section*{III. Non-adiabatic transport of atoms} 

%
%
%
Our proposal is based on the collisional interaction of atoms brought to the same lattice site, and high-fidelity transport of atoms through the lattice is therefore crucial.
The transport of a single  atom to an empty lattice site is illustrated in Fig.\,\ref{fig:setup}b). The optical tweezer is first focussed on the site containing the atom and its power is ramped up. Then its position is translated to the destination site and the power is ramped down.
Since collisional gates are sensitive to the vibrational state, it is crucial to avoid vibrational excitation in each step.
In the following we first review the free-space harmonic oscillator theory and subsequently numerically model the process in the presence of the lattice potential.
We show that dynamics during the ramp-up is essentially unchanged, whereas the fidelity of non-adiabatic transport is significantly reduced by the presence of the lattice potential. We then numerically optimize the parameters controlling the transport and thereby regain high fidelity transport.

\subsection*{A. Optical tweezer ramp-up}
In the harmonic oscillator approximation, the vibrational excitations during ramp-up of the optical tweezer power can be treated analytically~\cite{Zhang:2007}. Because the matrix elements of the corresponding Hamiltonian vanish between states with opposite parity, the change of the trapping frequency can only induce excitations to the second excited state. In the low-excitation limit, the time-dependent Schr\"odinger equation is then solved in the Hilbert space consisting of only the ground state $\ket{\phi_g}$ and the second excited state $\ket{\phi_e}$, separated by the energy gap $\Delta E_{ge}=2\hbar\omega$.

The adiabaticity criterion for a change in the trap frequency is
\begin{equation}
\label{eq:adiabaticity-rampup}
\hbar \left|\frac{\mathrm{d} \omega(t)}{\mathrm{dt}}\right|=\xi\frac{(\Delta E_{ge})^2}
{\left|\bra{\phi_e}\frac{\partial H}{\partial \omega}\ket{\phi_g}\right|}~,
\end{equation}
where $\left|\bra{\phi_e}\frac{\partial H}{\partial \omega}\ket{\phi_g}\right|=\hbar/\sqrt{2}$ with the Hamiltonian $H$ of an harmonic oscillator with varying trap frequency $\omega(t)$.
If we keep the adiabaticity factor $\xi\ll1$ constant, the above differential equation determines the shape of the power ramp-up. The total ramp-up time $\Tr$ from an initial trap frequency $\omega_o$ to a final trap frequency $\omega_f$ is related to the adibaticity factor via~\cite{Zhang:2007}
\begin{equation}
\label{eq:rampTime}
\Tr=\frac{1-\frac{\omega_o}{\omega_f}}{4\sqrt{2}\xi\omega_o},
\end{equation}
and the excitation probability is given by
\begin{eqnarray}
\label{eq:Pexc-rampup}
\mrm{P}{e}^{\rm harm}(t) &=& \mrm{P}{e}^0 \cdot\sin ^2\left[\frac{\sqrt{2 \xi ^2+\frac{1}{2}}\cdot \log\left[1-4\sqrt{2}t\xi\omega_o\right]}{4\xi}\right].
\end{eqnarray}
The excitation probability displays an oscillatory behavior with an envelope $\mrm{P}{e}^0=4\xi^2/(1+4\xi^2)$.
The oscillatory factor in Eq.\,(\ref{eq:Pexc-rampup}) predicts that with an appropriate timing, the ramp-up can be done non-adiabatically with in principle unity fidelity. The existence of minima in the non-adiabatic excitation probabilities have been demonstrated experimentally using a cloud of atoms in a dipole trap with a longitudinally translated focus~\cite{Couvert:2008}.

\begin{figure}[!t]
  \centering
  \includegraphics[width=0.9\columnwidth]{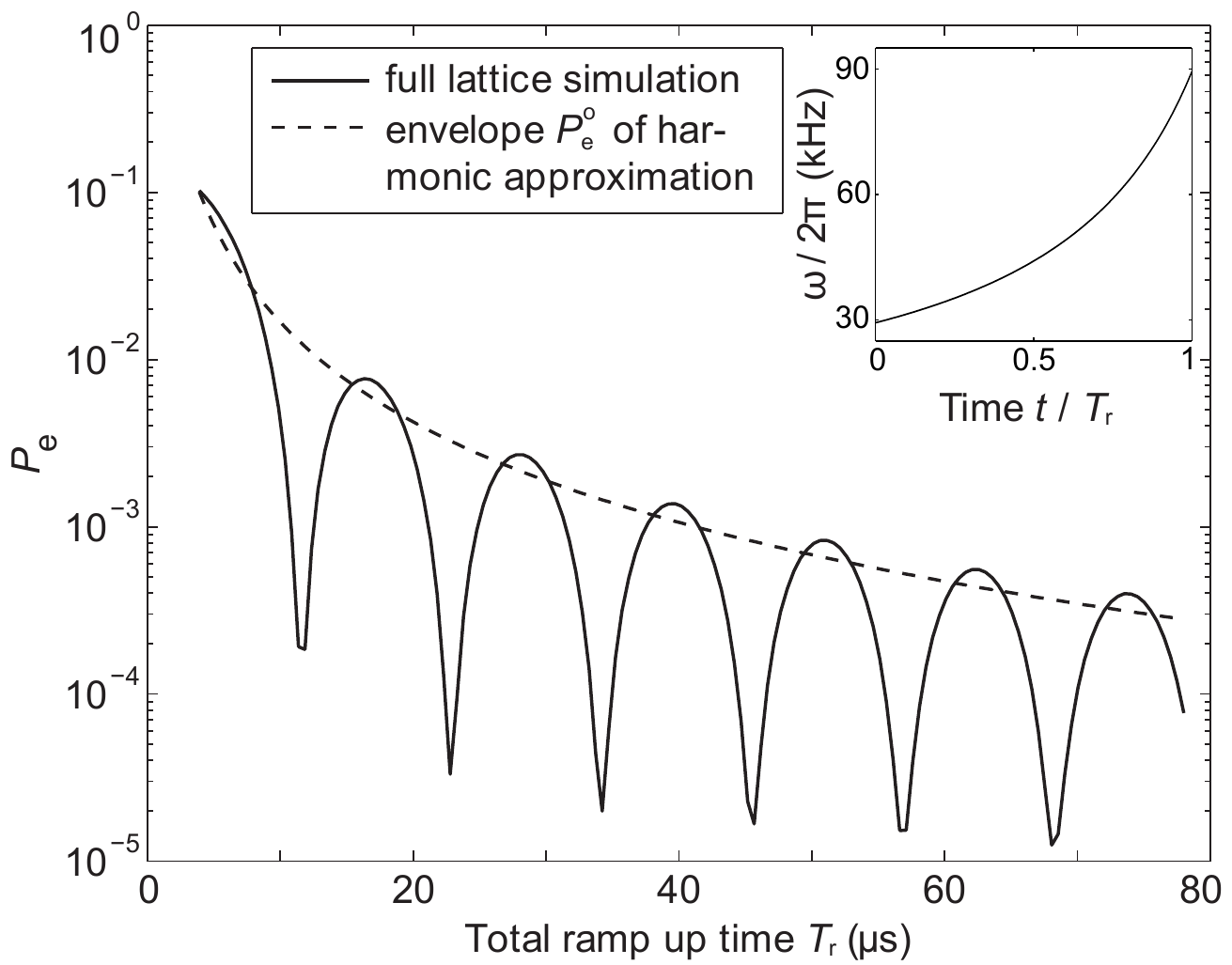} 
  \caption{Excitation probability versus the total ramp-up time $\Tr$
      of an optical tweezer centered on a lattice site to a depth of 500\,$\Erec$ according to Eq.\,(\ref{eq:adiabaticity-rampup}). The inset shows the trapping frequency of the potential well as a function of time.}
  \label{fig:rampup1}
\end{figure}
Since realistic lattice potential wells are not fully harmonic, we have studied the ramp-up of a Gaussian-shaped optical tweezer potential superimposed on a cosine lattice potential. We numerically solve the time-dependent Schr\"{o}dinger equation as the depth of the tweezer is increased from zero to 500 $\Erec$ according to the adiabaticity condition of Eq.\,(\ref{eq:adiabaticity-rampup}).

The resulting excitation probability $P_e^{\rm full}(\Tr)$ as a function of the total ramp-up time $\Tr$ (solid line) decreases for larger ramp-up times illustrating the transition from non-adiabatic to adiabatic dynamics (Fig.\,\ref{fig:rampup1}).
For comparison, we also show the envelope of the harmonic oscillator approximation $\mrm{P}{e}^0$ with $\xi$ calculated from the given total ramp time $\Tr$ via Eq.\,(\ref{eq:rampTime}) (dashed line in Fig.\,\ref{fig:rampup1}) and observe a decent but not exact agreement with the numerical result. The small disagreement is due to the anharmonicity of the lattice potential and we have verified that it diminishes as the optical tweezer is made tighter. Our calculation also shows that very small excitation errors can be reached with carefully chosen ramp timing.
For example, errors below $P_e^{\rm full}=10^{-3}$ are realized at the first minimum at a ramp-up time of $\Tr=11\,\mu$s.




\subsection*{B. Atom transport}
Atoms are transported in the lattice by shifting the position $x_o$ of the optical tweezer. Again, we are interested in the excitation probability during this process and we first discuss the transport of atoms in a harmonic potential without the presence of the lattice, which can be solved analytically. The effects of non-adiabaticity can be calculated by  replacing $\omega$ with $x_o$ in Eq.~(\ref{eq:adiabaticity-rampup}) and using $\Delta E_{ge}=\hbar\omega$, as $\frac{\partial H}{\partial x_o}$ shifts the parity of states. In the harmonic approximation $\left|\bra{\phi_e}\frac{\partial H}{\partial x_o}\ket{\phi_g}\right|=\hbar \omega/(\sqrt{2}\sigma_o)$, where $\sigma_o=\sqrt{\hbar/(m\omega)}$ is the harmonic oscillator length.

Again, we start with a translation profile which keeps the adiabaticity parameter $\xi$ constant. This requires a displacement with a constant velocity $v_o=\frac{\mathrm{d} x_o}{\mathrm{dt}}=\xi \sqrt{2} \sigma_o \omega$ and yields the excitation probability
\begin{equation}
\label{eq:Pexc-transport}
\mrm{P}{e}^{\rm{harm}}(t)= \mrm{P}{e}^0 \cdot \sin^2 \left[\sqrt{1+4 \xi ^2}\omega t/2\right],
\end{equation}
with $\mrm{P}{e}^0$ as defined above. To estimate the time for an adiabatic transport over one lattice site, we consider the adiabaticity parameter $\xi=0.016$ for which the envelope $P_e^{0}$ of the excitation probability has dropped to $10^{-3}$. For a potential depth of $500\,\Erec$ of the optical tweezer ($\mrm{\omega}{trap}\approx 2\pi\cdot90\,$kHz, $\sigma_o=36\,$nm), this yields a velocity of $v_o=0.45\,\mu$m/ms or a transport time of $T_t=v_0 \cdot \alat=1.2\,$ms over the distance of a single lattice spacing. 

\begin{figure}[!t]
  \centering
  \includegraphics[width=0.9\columnwidth]{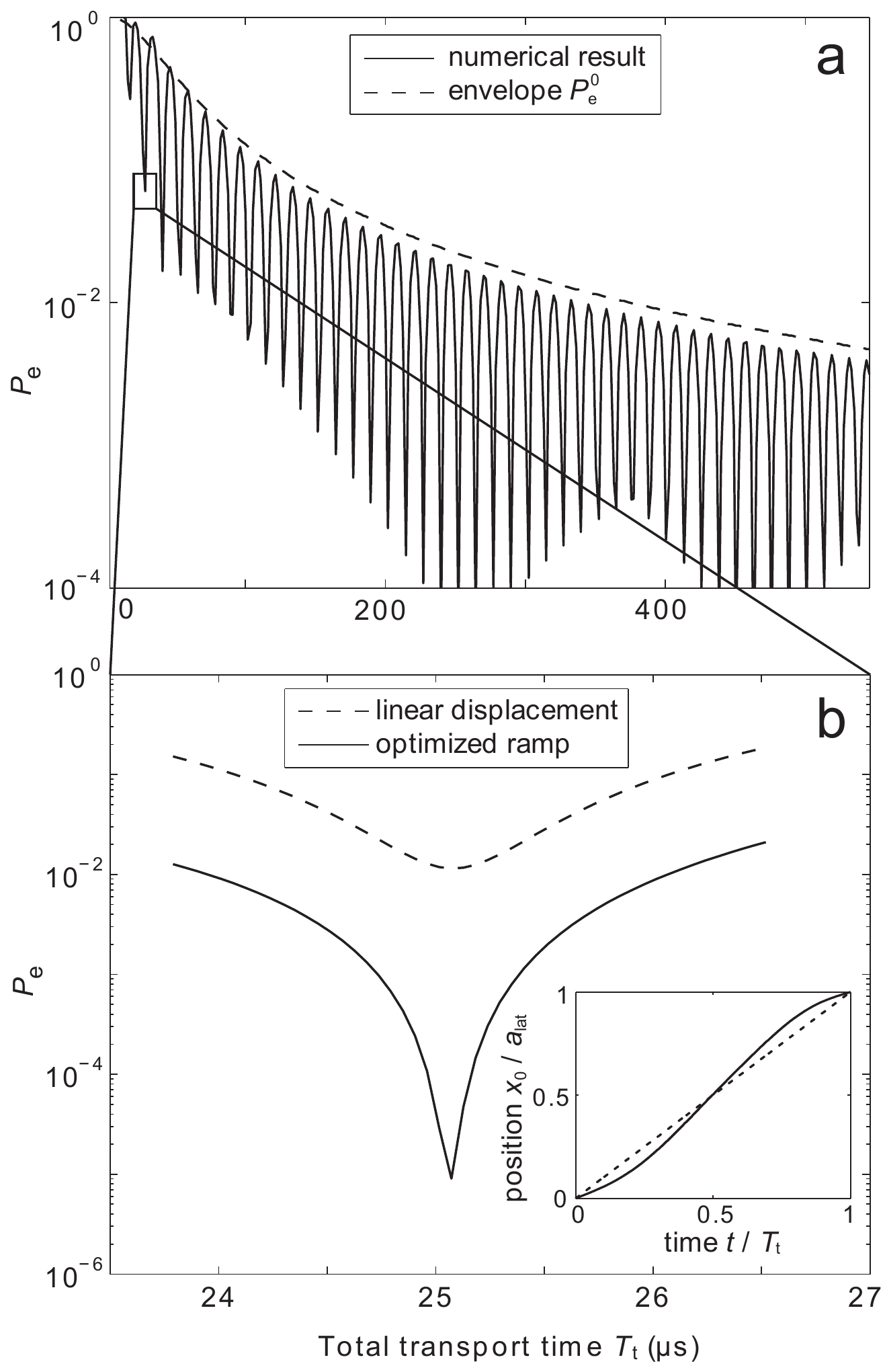} 
  \caption{Excitation probability versus transport time.
  a) Linear displacement of the optical tweezer. Numerical result $\mrm{P}{e}^{\rm full}$ for the excitation probability including the influence of the lattice and the profile of the tweezer (solid line) and envelope $\mrm{P}{e}^0$ of the result for the transport of an harmonic oscillator potential without a lattice (dashed line).
  b) Zoom into the second minimum in a). Numerical result for the linear displacement (dashed line) and an optimized ramp shape (solid line). The optimization reduces the excitation probability to below $10^{-4}$.
  The inset shows the displacement curve of the linear ramp (dotted line) and the optimized ramp (solid line) including harmonics up to the fifth order.}
  \label{fig:transport}
\end{figure}

We now include the lattice and the Gaussian profile of the optical tweezer and investigate to which extent non-adiabatic transport can be realized.
Fig.\,\ref{fig:transport}a) shows our numerical solution of the excitation probability $P_e^{\rm full}$ for a translation of the tweezer position over one lattice site at a constant speed as a function of the total transport time $T_t$.
Again, we observe an oscillatory behavior and a good agreement of the upper envelope curves. However, the numerical results do not reach zero excitation probability during the oscillations.
An enlarged view of one of the minima at short transport times shows that the excitation probability stays quite high (Fig.\,\ref{fig:transport}b)).

To improve this, we parameterize the translation profile as a linear ramp plus a number of harmonics and optimize their weights using a standard simplex algorithm~\cite{Nelder:1965}. Using only five harmonics dramatically improves the fidelity [solid curve in Fig.\,\ref{fig:transport}b)].
As can be seen in the inset, the translation profile is only slightly modified compared to the linear profile and the dominant component is a single harmonic with a period corresponding to the duration of the displacement.

We have investigated how the error accumulates as the qubit is transported for longer distances across multiple sites. We have simply repeated the optimal single-site displacement found above with $T_t=25\,\mu$s per lattice site up to 100 times. We find that the error after multi-site transport does not accumulate with the number of sites. Instead, it oscillates with a period of approximately eight lattice sites and is bound by a maximal error of only a few times the error of a transport over a single lattice site.
Thus, fast shutteling of atoms within the quantum register should be feasible with total excitation probabilities well below $10^{-3}$.

For the transport tweezer we would choose a wavelength red detuned compared to both excited state levels (D1 and D2 line) such that both qubit states are transported simultaneously. If the polarization is chosen to be linear, both states experience the same light shift and do not undergo a differential phase evolution (the residual differential light shift drops below $10^{-3}$ for a detuning of $\Delta=4\,$nm). The effect of the transport beam on the neighboring sites can be neglected~\cite{Zhang:2006}.

\section*{IV. Two-qubit gates}
\label{sec:2-qubit-gates}

In the section above we have described how we can position two selected atoms in neighboring sites as a starting point for a two-qubit gate. The gate can be accomplished by collisional interactions between atoms in the same well, and we will analyze three distinct proposals based on controlled atomic collisions, utilizing either the spin-dependent interaction, spin-dependent transport or spin-exchange interactions.

The quantum bit is encoded in a pair of atomic states, one from each hyperfine split ground state manifold in $\Rb$. We denote the upper and lower states $\ket{\uparrow}$ and $\ket{\downarrow}$, and in the following two subsections we show how a pair of atoms can acquire a phase evolution depending on the states of the two qubits,
$\ket{jk}\rightarrow e^{iU_{jk}t/\hbar}\ket{jk}$, for $j,k=\uparrow,\downarrow$, and how the interaction strengths $U_{jk}$ can be suitably controlled.


\subsection*{A. Merging two qubits into a single well} 

For the first two-qubit gate we discuss, one merges the two atoms into the ground state of a single combined well.
The atoms have the same wave function and the interaction energy is calculated from an integral over the fourth power of the wave function. For cold atoms, the short range interaction strength is proportional to the s-wave scattering length, and near a Feshbach resonance  there can be a difference between the scattering lengths for different and identical spin states.
For the states $\ket{F=1, m_F=+1}\equiv\ket{\uparrow}$ and $\ket{F=2, m_F=-1}\equiv\ket{\downarrow}$ of $\Rb$, the interaction strengths at a magnetic field of $B=9.12\,$G are given by
$U_{\uparrow\uparrow}\approx U_{\downarrow\downarrow}\approx0.9\,U_{\uparrow\downarrow}$~\cite{Widera:2008}.
Thus, by timing the merging and separation of the wells one can realize a phase gate.
For an optical tweezer of potential depth $500\Erec$, as considered here, a differential phase of $\pi$ between the two hyperfine states is acquired in roughly 0.8\,ms.

A major experimental limitation with this approach is that during the merging and splitting of the atoms, the atomic potential acquires the form of a double well with a sizable tunneling rate. The splitting must be performed slowly compared to the tunneling rate in order to ensure that one atom goes to each site. Note also that, in the absence of interaction between the atoms, one cannot map a doubly occupied state into two singly occupied orthogonal states. Thus, fast non-adiabatic merging and separation is not possible within this approach.



\subsection*{B. Gate using spin-dependent transport}
\begin{figure}[!t]
  \centering
  \includegraphics[width=0.9\columnwidth]{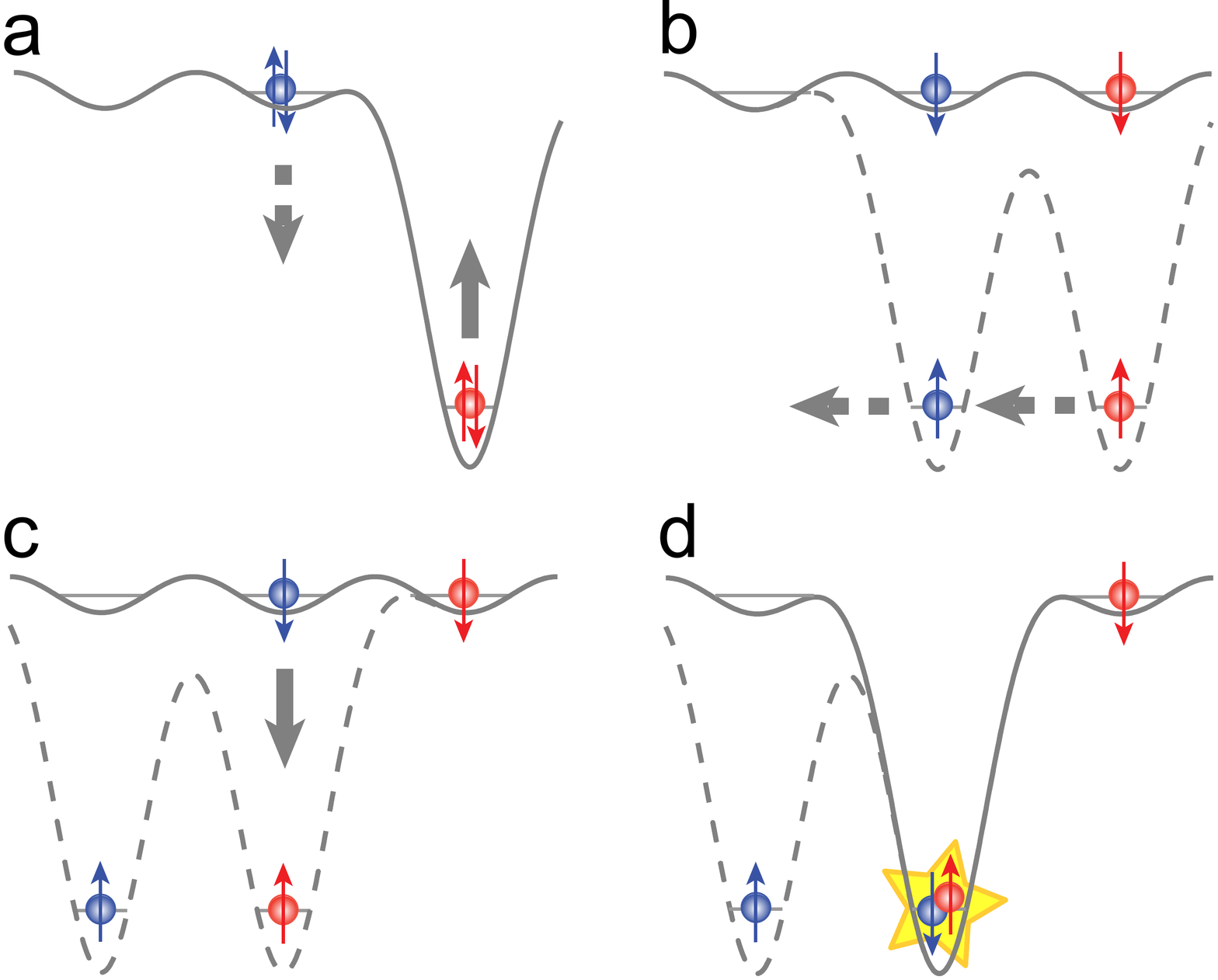} 
  \caption{Phase gate using spin-dependent transport.
  a) The two atoms (blue and red) are in an arbitrary superposition between $\ket{\downarrow}$ and $\ket{\uparrow}$ and are located at neighboring sites after the red atom has been transported. Here the tweezer is still at the full depth as used for the transport. The lattice site to the left of the pair is empty (Fig.\,\ref{fig:setup}).
  b) While the tweezer on the right atom is switched to circular polarization acting only on the $\ket{\uparrow}$ component, a second tweezer with circular polarization is ramped up at the position of the left atom.
  c) The $\ket{\uparrow}$ components of both atoms are moved one lattice site to the left.
  d) The polarization of the right tweezer is switched to linear, effectively ramping up the potential for the $\ket{\downarrow}$ component. The atoms in the central lattice site acquire a collisional phase $\phi$.
  Subsequently the steps are reversed to restore the original configuration.}
  \label{fig:phaseGateScheme}
\end{figure}

To realize a faster gate, we propose to adapt the spin-dependent lattice transport process~\cite{Mandel:2003} to our architecture
by utilizing spin-dependent light shifts in such a way that only particular spin combinations are merged. By using  $\ket{F=2,m_F=-2}\equiv\ket{\uparrow}$ and $\ket{F=1,m_F=-1}\equiv\ket{\downarrow}$ and an appropriate laser detuning between the two excited state fine structure levels ($420.86~\mathrm{nm}$) the light shift for the state $\ket{\downarrow}$ cancels when using $\sigma^-$ polarized light.
\footnote{Also, using $\sigma^+$ polarized light, one can cancel the light shift of either of the states, but the ratio of light shift and scattering rate is worse in this case.}
We propose to perform a quantum phase gate in five steps (Fig.\,\ref{fig:phaseGateScheme}).

We consider an atom in a superposition state of $\ket{\uparrow}$ and $\ket{\downarrow}$. First, the $\ket{\uparrow}$ part of the wave function is moved to an empty row. Next, the $\ket{\uparrow}$ part of a neighboring qubit is moved to overlap with the remaining $\ket{\downarrow}$ of the first atom. In this way, a controlled collision between these two spin states with an energy $U_{\uparrow\downarrow}$ takes place, while the other combinations of spin states undergo no interaction. The spin state thus acquires a phase shift $\phi$, which attains $\pi$ after a certain amount of time. Finally, the steps are reversed to restore the original configuration. Since each spin state resides in its own trapping potential during the collision, the two atoms can be rapidly separated after the collisional interaction.

Unlike in the transport discussed in previous sections, the phase shift during merging and interaction is acquired by only one of the two-qubit spin state combinations. Since the phase depends deterministically on the intensity profile during transport
and on the transport time,  it can be calculated and corrected for. Choosing intensity and transport profiles where the light shift is symmetrical with respect to the center of the pulse sequence, a robust compensation can also be achieved by applying a $\pi$-pulse after half of the sequence
to reverse the phase evolution and another $\pi$-pulse at the end to
restore the correct populations.

To optimize the speed of the collisional phase gate, we switch the optical tweezer to linear polarization after the merging to ramp up the tweezer for the other spin state and enhance the interaction strength. For a tweezer depth of  $500\,\Erec$ the interaction energy is $U=h\cdot 6\,$kHz and a phase shift of $\pi$ is acquired in $t_\pi\approx 83\,\mu$s. In Table~\ref{tab:transportGate} we summarize the times required for the individual steps, obtaining $t_\mathrm{gate}\approx(200+n \cdot 50)\mu$s, where $n$ is the transport distance in units of lattice sites. In a possible implementation, the transport of several atoms could be performed in parallel to keep the typical values of $n$ per gate low.
Although this does not improve the individual gate error, it does decrease the effective time per gate and thus diminishes the effect of background decoherence effects.


%
%

\begin{table}
    \begin{center}
    \begin{tabular}{ | c | c | c |}
    \hline
    step & amount & time\\ \hline\hline
    ramp-up/down & 6 & $11\,\mu$s \\ \hline
    transport & $2(n+1)$ & $25\,\mu$s \\ \hline
    phase gate & 1 & $83\,\mu$s \\ \hline \hline
    overall & - & $(199+n 50)\,\mu$s\\ \hline
    \end{tabular}
    \caption{Time budget for the spin-dependent transport gate involving a transport over $n$ lattice sites.}
    \label{tab:transportGate}
    \end{center}
\end{table}

The figure of merit for the speed of a quantum gate is the ratio of the gate time to the coherence time of the qubit. The  hyperfine states which allow spin-dependent potentials, such as $\ket{F=1,m_F=-1}$ and $\ket{F=2,m_F=-2}$ are sensitive to magnetic fields and it is therefore crucial to reduce the magnetic field noise to obtain reasonable coherence times. The dephasing time $T_c=h/\Delta E$ of the two states is given by their relative energy shift, $\Delta E=3/2 \mu_B B=h \cdot 2.1\,$kHz/mG. 
It is state of the art of active magnetic field stabilization to reach a short term stability at sub-milligauss level ~\cite{Zibold:2007,Gross:2010}. A magnetic field noise of $50\,\mu$G yields a coherence time of $T_c=10\,$ms and thus leads to a phase error of $10^{-2}$ within the gate time of $T_g\sim100\,\mu$s. An improvement to $5\,\mu$G is extremely challenging but could be realizable with existing technology~\cite{Laurent:2006} and it would lead to an error of $10^{-3}$.



\subsection*{C. Spin-exchange gate}

When using the clock states $\ket{F=1,m_F=-1}$ and $\ket{F=2,m_F=+1}$ of  $\Rb$
as qubit states, coherence times of many seconds have been observed working at the 'magic' magnetic field of 3.229\,G~\cite{Deutsch:2010}. In the following we present a gate proposal based on atoms in these clock states.
Spin-dependent potentials do not exist for the clock states, and we thus need to merge the atoms to the same potential. In contrast to the simple merging gate of Section\,IV.A, we propose to transfer the two qubits into different vibrational states of the combined potential instead of merging them into the ground state. This separation in orthogonal spatial wave functions will allow more robust merging and separation of the two qubits.

It can be done by mapping the left well qubit into the excited vibrational level of the combined well and the right well qubit into the ground vibrational level, while preserving the qubit state~\cite{Anderlini:2007}.
This mapping is described by:
\begin{equation}
\begin{array}{ccc}
  \alpha\ket{\uparrow}_\mathrm{L}+\beta\ket{\downarrow}_\mathrm{L} & \to &
  \alpha\ket{\uparrow}_\mathrm{e}+\beta\ket{\downarrow}_\mathrm{e} \\
  \tilde{\alpha}\ket{\uparrow}_\mathrm{R}+\tilde{\beta}\ket{\downarrow}_\mathrm{R} & \to &
  \tilde{\alpha}\ket{\uparrow}_\mathrm{g}+\tilde{\beta}\ket{\downarrow}_\mathrm{g},
\end{array}
\end{equation}
where $\ket{\cdot}_\mathrm{L}$ and $\ket{\cdot}_\mathrm{R}$ designate the wave function of the left and right potential well and $\ket{\cdot}_\mathrm{g}$, $\ket{\cdot}_\mathrm{e}$ are the ground and first excited vibrational states in the right potential well.

In the combined system the new eigenenergy basis is now formed by the singlet and the triplet states
\begin{equation}
\begin{array}{ccl}
  \ket{s} & = & \ket{\uparrow}_\mathrm{g}\ket{\downarrow}_\mathrm{e}-
  \ket{\downarrow}_\mathrm{g}\ket{\uparrow}_\mathrm{e}\\
  \ket{t_0} & = &
  \ket{\uparrow}_\mathrm{g}\ket{\downarrow}_\mathrm{e}+
  \ket{\downarrow}_\mathrm{g}\ket{\uparrow}_\mathrm{e}\\
  \ket{t_{-1}} & = &
  \ket{\downarrow}_\mathrm{g}\ket{\downarrow}_\mathrm{e}\\
  \ket{t_{+1}} & = &
  \ket{\uparrow}_\mathrm{g}\ket{\uparrow}_\mathrm{e}~.\end{array}
\end{equation}

For the $\sqrt{\rm swap}$ gate, we exploit the differential phase evolution between the singlet and the triplet states which stems from their different symmetry. As the singlet spin state $\ket{s}$ is antisymmetric, the spatial wave function of bosonic particles must also be antisymmetric, while the symmetric triplet state $\ket{t}$ leads to a symmetric wave function.
In the antisymmetric wave function of $\ket{s}$, the two atoms have essentially zero overlap, such that interactions play no role. In the symmetric wave function of state  $\ket{t_0}$, however, the collisional interaction energy has a finite value $U_{eg}$ and a corresponding phase is obtained. The value of $U_{eg}$ is related to the interaction energy $U_{gg}$ of two atoms in the ground state via integrals of the appropriate spatial distributions, and for harmonic oscillator states $U_{eg}=0.35\cdot U_{gg}$~\cite{Sherson:2010b}.

The differential phase evolution between $\ket{s}$ and $\ket{t_0}$ results in spin-exchange oscillations between $\ket{\uparrow}_\mathrm{g}\ket{\downarrow}_\mathrm{e}$ and $\ket{\downarrow}_\mathrm{g}\ket{\uparrow}_\mathrm{e}$:
\begin{equation}
\begin{array}{rl}
  \Psi(t=0) &= \ket{s}+\ket{t_0} \sim \ket{\uparrow}_g \ket{\downarrow}_e\\
  \Psi(t) &= \ket{s}+ e^{i U_{eg} t/\hbar} \ket{t_0}\\
  \Psi(t=T_{\rm swap}) &= \ket{s}-\ket{t_0} \sim \ket{\downarrow}_g \ket{\uparrow}_e\\
  \Psi(t=T_{\rm swap}/2) &= \ket{s}+i\ket{t_0} \sim \ket{\uparrow}_g \ket{\downarrow}_e+i\ket{\downarrow}_g \ket{\uparrow}_e~.\\
  \end{array}
\end{equation}
After a  time $T_{\rm swap}=\pi\hbar/U_{eg}$~\cite{Anderlini:2007} the states are swapped, while after $t=T_{\rm swap}/2$ an entangling $\sqrt{\rm swap}$ gate is realized.

\begin{figure}[!t]
  \centering
  \includegraphics[width=0.9\columnwidth]{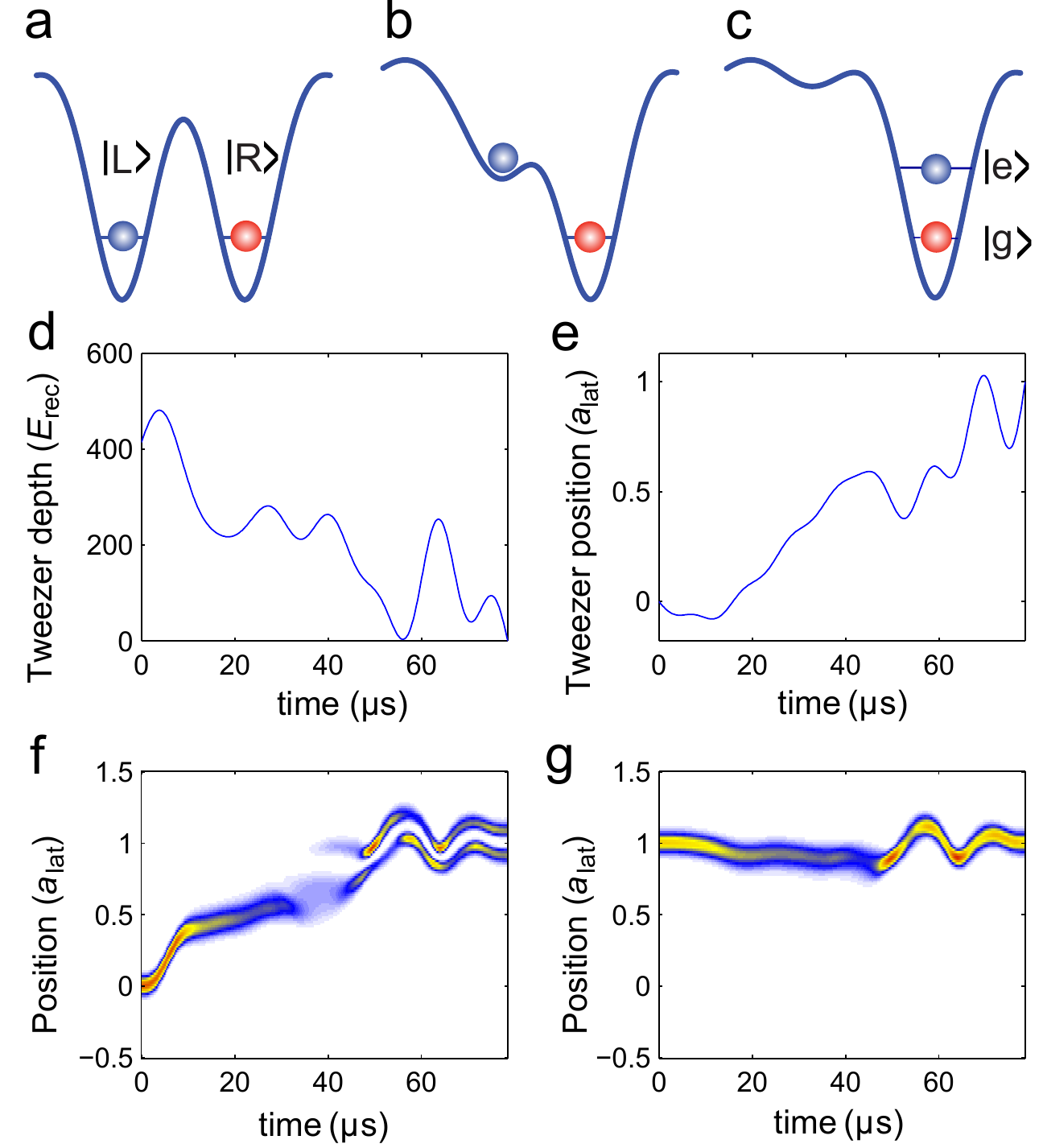} 
  \caption{Band mapping for the exchange gate.
  a)-c) Schematics of the potential formed by the lattice and the two tweezers at three different times. While the transport tweezer is moved from the left to the right lattice site and simultaneously ramped down, the atom in the left well is mapped to the first excited state in the right site.
  d) and e) Depth and position of the transport tweezer as a function of time for the optimized ramp shapes. The auxiliary tweezer is kept at a fixed depth and position (not shown).
  f) and g) Density profile versus time for the transported atom f) and the stationary atom g). At the end of the sequence the atom starting in the left well is mapped to the first excited state and has a wave function with a node, while the atom starting in the right well is unchanged.}
  \label{fig:bandmap}
\end{figure}

For the band mapping process, the two atoms for the two-qubit gate are transported to neighboring lattice sites [see Fig.\,\ref{fig:bandmap}a)-c)]. The transport tweezer on the left atom starts at a depth of $V_{\rm start}=400\,\Erec$ and is linearly ramped down to zero in $75\,\mu$s while it is simultaneously moved to the right lattice site. We find that the performance of the band mapping is improved if we apply a second auxiliary tweezer on the right lattice site, which is kept at a constant depth of $V_{\rm aux}=200\,\Erec$. This configuration leads to a mapping of the left atom to the first band on the right lattice site [Fig.\,\ref{fig:bandmap}a)-c)].

To optimize the fidelity of the band mapping, we refine the linear ramp shapes by adding harmonics of the ramp time to both the intensity and the position ramp of the transport tweezer. We solve the time-dependent Schr\"odinger equation and obtain the single-particle mapping fidelities as the overlap of the calculated wave function with the target states. We define the fidelity $F$ of the merging process as the product of the single-particle mapping fidelities of the two atoms.
Fig.\,\ref{fig:bandmap}d)-g) shows the resulting ramps and wave functions for an optimization including harmonics up to  15th order corresponding to a spatial period of $0.13\,\alat$, for which we obtain an infidelity of $1-F=2\cdot 10^{-4}$.

A similar analysis for double-well lattices resulted in a saturation of the merging error of the order of $10^{-2}$~\cite{DeChiara:2008}, which is significantly higher than our results. We attribute this to the difference in available control parameters\footnote{In Ref.~\cite{DeChiara:2008} the control parameters are the overall power of the double well lattice, the ratio of horizontal to vertical power, and the phase between the two, which indirectly determine the position and relative depths of the two sides of the double well.}.
We have also included interactions into the simulations but the transfer is so fast that the effect on the overall band mapping fidelity is negligible. It is  important to  calculate the accumulated interaction phase during the band mapping as it contributes to the total gate phase. In the case illustrated in Fig.\,\ref{fig:bandmap}, the accumulated interaction phase during the mapping was $0.15\,$rad.

After the band mapping, we wait for a time $T_{\rm swap}/2$ to perform the $\sqrt{\rm swap}$ gate before reversing the band mapping process. For comparison with the gate proposal discussed in Sec.\,IV~B
we introduce an $11\,\mu$s ramp up of the tweezer to $500\,\Erec$, which results in a reduction of the swap time to $T_{\rm swap}/2=125\,\mu$s.


In summary, the exchange gate requires four ramps of the tweezer power (each $\sim 11\,\mu$s), the transport ($25\,\mu$/site), the merging and splitting of the wells ($150\,\mu$s), and finally the phase gate ($125\,\mu$s) (see Tab.\,\ref{tab:exchangeGate}). In total this gives a gate time $t_\mathrm{gate}\approx(300+n \cdot 50)\,\mu$s, where $n$ is the distance to the interaction site. The gate time is therefore more than $10^4$ times shorter than the decoherence time of the clock states.


%

\begin{table}
    \begin{center}
    \begin{tabular}{ | c | c | c |}
    \hline
    step & amount & time\\ \hline\hline
    ramp-up/down & 2 & $11\,\mu$s \\ \hline
    transport & $2\cdot n$ & $25\,\mu$s \\ \hline
    phase gate & 1 & $125\,\mu$s \\ \hline
    merge/split & 2 & $75\,\mu$s \\ \hline \hline
    overall & - & $(297+n\cdot 50)\,\mu$s\\ \hline
    \end{tabular}
    \caption{Time budget for the exchange gate.}
    \label{tab:exchangeGate}
    \end{center}
\end{table}

\section*{V. Discussion of errors}

In the following we discuss the contribution of error and decoherence sources other than the vibrational excitations considered so far.

\subsection*{A. Spontaneous scattering}

A fundamental source of error is the spontaneous scattering due to the off-resonant absorption of lattice and optical tweezer photons. Due to the large detuning, the absorption of lattice photons can be neglected. With a scattering rate of $\Gamma_{sc}\approx 4\cdot10^{-2}\,$Hz for all three lattice axes, the probability to scatter a photon during one gate time is $<10^{-6}$. 

For the optical tweezer, the spontaneous emission can be suppressed by choosing a large detuning. A detuning of 10\,nm requires a laser power of 1\,mW to reach the $500\,\Erec$ potential depth. This yields a scattering rate of $\Gamma_{\rm sc}=0.1\,$Hz and a scattering probability during one gate time $T_g\sim300\,\mu$s of $P_{\rm sc}=\Gamma_{\rm sc} T_g=3\cdot10^{-5}$.


In the case of spin-dependent transport, the detuning is much smaller since the wavelength of the optical tweezer has to lie between the two fine structure lines. At the chosen waist radius, the potential depth of $500\,\Erec$ is reached at a power of $70\,\mu$W yielding a spontaneous scattering rate of $\Gamma_{\rm sc}=1.5\,$Hz and a scattering probability of $P_{\rm sc}=4\cdot10^{-5}$ during one transport time.


\subsection*{B. Pointing stability and intensity noise}

\begin{figure}[!t]
  \centering
  \includegraphics[width=0.9\columnwidth]{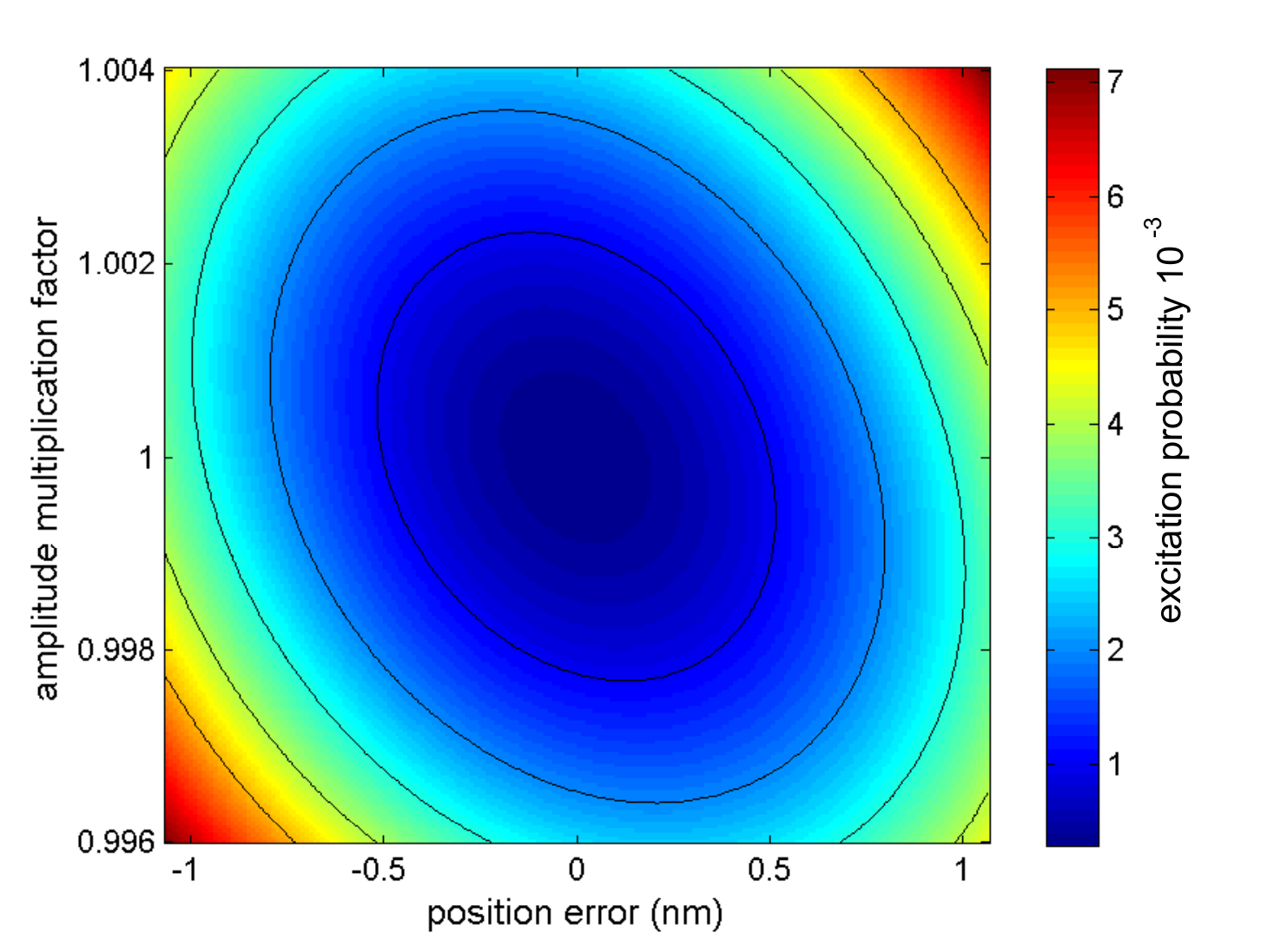} 
  \caption{Band-mapping infidelity as a function of intensity noise of position errors. For the transport tweezer we introduce a positioning error and a scaling error in the amplitude. The contour lines show the multiples of an error of $10^{-3}$.}
  \label{fig:bandmapError}
\end{figure}

We now discuss the influence of experimental imperfections on the gate fidelity. For the present proposal the limitations are the pointing error and intensity fluctuations of the tweezer. In order to investigate these effects quantitatively, we assume only  low-frequency noise due to shot-to-shot fluctuations. We calculated the gate infidelity in the presence of a constant offset in the pointing position and of a reduction or increase of the optical tweezer intensity by a constant factor.
In particular, we included these errors into the simulation of the band mapping process discussed in Sec. IV~C.
As can be seen in Fig.\,\ref{fig:bandmapError}, both parameters have to be very well controlled in order to avoid a significant reduction of the fidelity. In order to keep the gate infidelity below $10^{-3}$, our scheme requires a relative intensity stability of $10^{-3}$ and pointing precision in the nm range. This intensity stability is definitely manageable whereas the required pointing stability is beyond the current state-of-the-art experiments, which achieve a pointing error of 50\,nm~\cite{Weitenberg:2011}, where full active pointing stabilization was, however, not implemented.
It might be possible to find conditions under which the performance is less sensitive to errors in the input controls as it was  done for transport of harmonic traps~\cite{Murphy:2009}.

Intensity noise causes additional dephasing in the spin-dependent two-qubit gate of Section\,V.B, since the depth of  $500\,\Erec=h\cdot 1\,$MHz implies a phase accumulation of several tens of $2\pi\,$rad during the interaction time. To keep the absolute dephasing error below $2\pi \cdot 10^{-3}\,$rad, the relative intensity noise has to be below $10^{-5}$. This is technically challenging but is possible as demonstrated in the LIGO project~\cite{Nocera:2004}, where the intensity noise is reduced to a fractional level of $10^{-8}$ over the relevant spectral band.


\subsection*{C. Error threshold for fault-tolerance}
We have argued that combined transport and gate errors at the $10^{-3}$ level can be reached with our proposal. With errors of this magnitude, error correcting codes may be applied to reach scalable fault tolerance, but only with a vast overhead in number of physical qubits and operations~\cite{Steane:2003}.
An alternative approach to fault tolerance uses stabilizer codes in a scheme where errors are not corrected, but error syndromes are forwarded to higher algorithmic levels, and here gate errors at the few percent level can be tolerated~\cite{Knill:2005}.
Fault tolerant quantum computing is a very active field of research, and very recent simulations show tolerance to percent level errors for a two-dimensional qubit structure on a lattice with only nearest neighbor interactions~\cite{Wang:2011}.
With our ability to move atoms around with low error, a simpler scheme with the same error threshold may well be possible.

\section*{VI. Conclusion}
In conclusion, we have proposed a complete architecture for scalable quantum computing based on transport of atoms in an optical lattice by movable optical tweezers. We have presented explicit calculations of the errors occurring in the different processes of our proposals, and we have numerically optimized non-adiabatic ramps of the tweezer potentials to a vibrational excitation error below $10^{-3}$. Our analyses show that ultracold atoms in optical lattices are indeed a promising candidate for fault-tolerant scalable quantum computing.

The authors thank Immanuel Bloch, Nicolaj Nygaard, and Uffe V. Poulsen for fruitful discussions and J.F.S. acknowledges  support from the Danish Council for Independent Research $|$ Natural Sciences and EU (Marie Curie Individual Fellowship).



\bibliographystyle{prsty}
\bibliography{References}

\end{document}